\documentclass[10pts,twocolumn,romanappendices]{IEEEtran}

\usepackage{cite,calc}
\usepackage{graphicx,colordvi,psfrag,cite}
\usepackage{amsmath,amssymb}
\usepackage{epstopdf}
\usepackage{epsfig}
\usepackage{calc,pstricks, pgf, xcolor}
\makeatletter
\newif\if@restonecol
\makeatother

\usepackage{algorithm2e}
\IEEEoverridecommandlockouts

\newcommand{\snr}{\text{$\mathsf{SNR}$}}

\newcommand{\bx}{\mathbf{x}}

\newtheorem{proposition}{Proposition}
\newtheorem{theorem}{Theorem}

\newtheorem{definition}{Definition}
\newtheorem{lemma}{Lemma}

\newcommand{\modulo}{\mbox{\,mod\,}}
\newcommand{\ISI}{{\rm ISI}}
\begin{document}

\title{Cyclic-Coded Integer-Forcing Equalization}
\author{Or~Ordentlich  and
        Uri~Erez,~\IEEEmembership{Member,~IEEE}
\IEEEcompsocitemizethanks{
\IEEEcompsocthanksitem
This work was supported in part by the Israel Science Foundation under grant 1557/10 and  by the Binational Science Foundation under grant 2008455.   The material in this paper was presented in part at the Forty-Eighth Annual Allerton Conference on Communication, Control, and Computing, Monticello, Illinois, Sept. 29 –- Oct. 1, 2010.
\IEEEcompsocthanksitem
Or Ordentlich and Uri Erez are with the Department
of Electrical Engineering-Systems, Tel Aviv University, Ramat Aviv 69978, Israel.
\protect\\
E-mail: \{ordent,uri\}@eng.tau.ac.il.
}}

\maketitle
\begin{abstract}
A discrete-time intersymbol interference channel with additive Gaussian noise is considered, where only the receiver has knowledge of
the channel impulse response. An approach for combining decision-feedback equalization with channel coding is proposed, where decoding precedes the removal of intersymbol interference. This is accomplished by combining the recently proposed integer-forcing equalization approach with cyclic block codes. The channel impulse response is linearly equalized to an integer-valued response. This is then utilized by leveraging the property that a cyclic code is closed under (cyclic) integer-valued convolution. Explicit bounds on the performance of the proposed scheme are also derived.
\end{abstract}
\section{Introduction}
The intersymbol interference (ISI) channel with additive Gaussian noise is one of the most basic channel models arising in digital communications. Thus, considerable effort has been devoted to developing effective transmission schemes for this channel; see, e.g., \cite{Forney98}, for a comprehensive survey.
The channel is described by
\begin{align}
y_k & = x_k+\sum_{m\neq0} h_m x_{k-m} + n_k \nonumber\\
& =x_k+{\rm ISI}_k+n_k,
\label{eq2b}
\end{align}
where $\ISI_k$ is the ISI resulting from other data symbols, and  $n_k$ is additive white Gaussian noise (AWGN) with zero mean and unit power.

The channel model may further be characterized by the availability of channel state information (CSI), where we distinguish between the case where CSI is available to the transmitter and the receiver alike and the
case where CSI is available to the receiver only.
As we next briefly recall, while the distinction between these two cases does not make a great difference at high signal-to-noise ratios (SNR) (which is the main focus of this paper) in terms of capacity \cite{Lee&Mess} (i.e., whether water-filling may be performed or not), it is of significant consequence for the design and implementation of equalization and coding schemes.

In the past decades coding for AWGN channels has reached an advanced state, and practical coding schemes (e.g., turbo and LDPC codes) operating near capacity are known. It is thus desirable to combine AWGN coding and decoding techniques with equalization in a \emph{modular} way, with the aim of approaching the capacity of the ISI channel.

The multitude of approaches developed to achieve reliable communication over the ISI channel may be roughly divided into two classes: multi-carrier approaches and single-carrier approaches. Both approaches may in principle be used to approach the capacity of the ISI channel, but offer different practical tradeoffs as we briefly touch upon next.

In multi-carrier transmission, the ISI channel is transformed into a set of parallel AWGN subchannels, each subchannel corresponding to a different frequency bin and experiencing a different SNR. This approach has the advantage that the subchannels are (virtually) ISI free, and thus the problems of equalization and decoding are decoupled. However, it has some drawbacks: the alphabet size of the transmitted symbols is considerably enlarged, which in turn makes the approach inapplicable to some media, such as magnetic recording channels. A related phenomena associated with multi-carrier transmission is that it results in a high peak-to-average power ratio which may also be undesirable (see e.g., \cite{LitsynBook,PAPR}). Furthermore, when CSI is available only at the receiver, bit allocation is precluded, and channel coding and decoding become more difficult, due to the variation of the SNR across subchannels.

Single-carrier approaches try to eliminate  most of the ISI without severely increasing noise power. The simplest approach is linear equalization
consisting only of a ``feed-forward" equalizer (FFE), which roughly transforms the channel into an additive colored Gaussian noise channel, where the minimum mean-squared error (MMSE) criterion corresponds to (linearly) maximizing the signal-to-interference-plus-noise (SINR) at the ``slicer". 
Performance may be improved using (non-linear) decision-feedback equalization, in addition to FFE. Specifically, MMSE decision-feedback equalization (DFE) will be discussed in greater detail in the next section. In fact, as shown by Guess and Varanasi \cite{GuessVaranasiIT}, the MMSE-DFE architecture is optimal in the sense of attaining mutual information, and allows to approach capacity with AWGN encoding/decoding, if decisions (fed to the DFE) are based on codewords rather than symbols. See also \cite{Cioffi95,Forneyallerton04}.

Unfortunately, the Guess-Varanasi approach, while quite pleasing from a theoretical perspective, requires long interleaving as well as long zero-padding, which in turn requires long latency.
This drawback can be avoided if CSI is available at the transmitter by Tomlinson-Harashima precoding (\cite{Tomlinson,Lee&Mess}), which essentially moves the DFE to the transmitter, but is inapplicable if the transmitter has no knowledge of the channel.

The approach proposed in this paper allows to avoid error propagation {\em{without}} incorporating an interleaver and with CSI available at the receiver only. In essence, the proposed method enables to perform (soft or hard) block decoding {\em before} decision feedback is performed.

We build on the integer-forcing (IF) equalization approach, which was recently proposed \cite{IFmimo} in the context of general MIMO channels with CSI available at receiver only. In this approach, multiple streams are encoded using an identical linear code and the receiver equalizes the channel matrix to
any full-rank integer matrix rather than to the identity.

In the context of IF equalization, an ISI channel may be viewed as a Toeplitz matrix. This special structure allows to further replace
the multiple codewords (in the context of ISI, this number would be large) with a {\em single} codeword provided that the linear code is further
a {\em cyclic code}.\footnote{The crucial element needed is that it be a linear shift invariant code, not necessarily cyclic.}

The paper is organized as follows. Section~\ref{sec:IFEQ} recalls some basic results on single-carrier equalization.
Section~\ref{sec:2} describes IF equalization as well as the general structure of the proposed scheme.  Section~\ref{sec:3}
derives the criteria for choosing the FFE equalizer. Section~\ref{sec:4} derives bounds on the attained performance. In Section~\ref{sec:performance}, the performance of the scheme is analyzed, and some examples are given.
Section~\ref{sec:6a} discusses practical coding techniques for integer-forcing equalization at high transmission rates.
The paper concludes with Section~\ref{sec:7}.
\section{Preliminaries}
\label{sec:IFEQ}
\begin{figure*}[htb]
\psfrag{input}{ $x_k$}
\psfrag{output}{ $y_k$}
\psfrag{noise}{ $n_k$}
\psfrag{minus}{\tiny  $-$}
\psfrag{m}{ $m$}
\psfrag{mhat}{ $\hat{m}$}
\psfrag{H}{$H(D)$}
\psfrag{A}{$A(D)$}
\psfrag{S}{\small $\Sigma$}
\psfrag{inputhat}{ $\hat{x}_k$}
\psfrag{output2}{ $y'_k$}
\psfrag{y2}{ $y''_k$}
\psfrag{G}{ $G(D)$}
 \includegraphics[width=2\columnwidth]{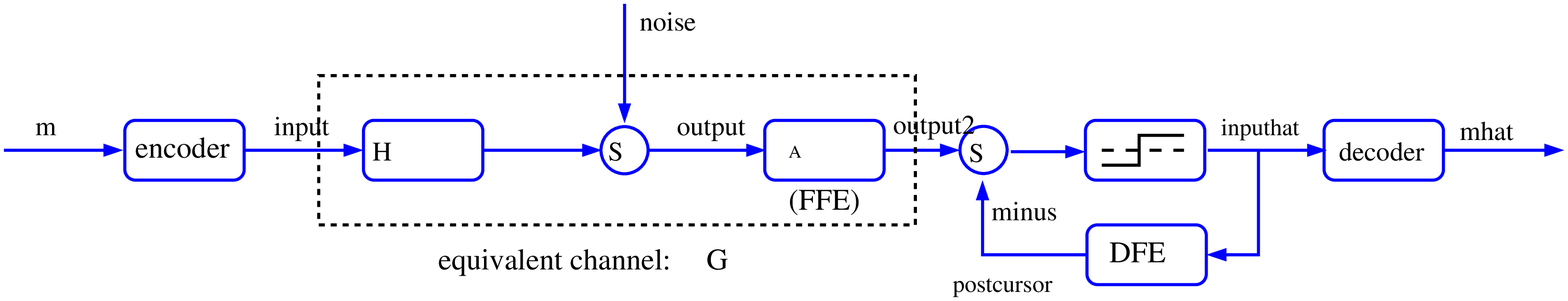}
\psfrag{input}{$x_k$}
\psfrag{output}{$y_k$}
\psfrag{noise}{$z_k$}
\psfrag{minus}{\tiny  $-$}
\psfrag{m}{$m$}
\psfrag{mhat}{$\hat{m}$}
\psfrag{H}{$G(D)$}
\psfrag{S}{\small $\Sigma$}
\psfrag{inputhat}{$\hat{x}_k$}
\psfrag{output2}{$y'_k$}
\psfrag{GF}{$\ \ \  \ \text{DFE}$}
 \includegraphics[width=2\columnwidth]{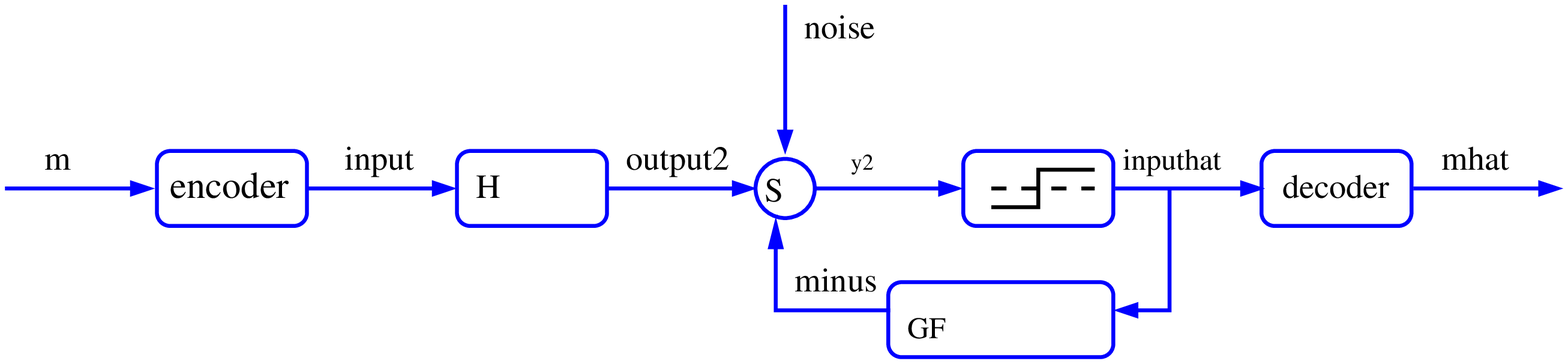}
 \caption{Decision-feedback equalization.}
 \label{fig:DFE}
 \psfrag{input}{$x_k$}
\psfrag{output}{$y_k$}
\psfrag{noise}{$n_k$}
\psfrag{m}{$m$}
\psfrag{S}{\small $\Sigma$}
\psfrag{minus}{\tiny $-$}
\psfrag{mhat}{$\hat{x}_k$,$\hat{m}$}
\psfrag{H}{$H(D)$}
\psfrag{inputhat}{$\hat{x}_k,\hat{m}$}
\psfrag{inputtilde}{$\hat{x}'_k$}
\psfrag{output2}{$y'_k$}
\psfrag{FB}{$I(D)-1$}
\psfrag{IF}{$\frac{I(D)}{H(D)}$}
\psfrag{mod}{\small$\bmod q$}
 \includegraphics[width=2\columnwidth]{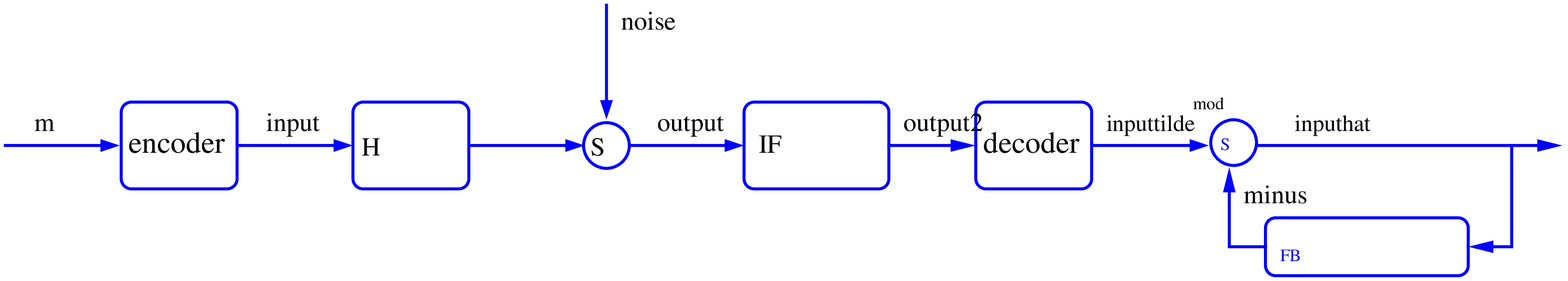}
\caption{A schematic description of an IF-DFE system.}
 \label{fig:if}
\end{figure*}

We briefly review basic single-carrier equalization architectures.
In the sequel, we use $D$-transform notation for sequences; i.e., a sequence
$\{ s_k \}$ is represented by its $D$-transform $S(D) = \sum_k s_k D^k$.
For example the channel (\ref{eq2b}) may be expressed as,
$$Y(D)=H(D)X(D)+N(D).$$
The simplest criterion for linear equalization (LE) is that of Zero-Forcing (ZF), where the ISI is completely canceled using an FFE only. This corresponds to taking the front end  (linear) filter to be $${A_{\text{ZF}}(D)}=\frac{1}{{H(D)}},$$
resulting in the equalized channel response $G(D)=1$. The induced noise enhancement can be large, especially when ${H(D)}$ has zeros near the unit circle. A variant that takes into account both ISI and noise enhancement is the linear MMSE equalizer $${A_{\text{MMSE-LE}}(D)}=\frac{H^*(D^{-*})}{{H(D)H^*(D^{-*})+1/\text{SNR}}}.$$
The MMSE-LE suffers from smaller (and in particular bounded) noise enhancement while allowing some residual ISI. The MMSE criterion is equivalent to maximizing the SINR at the slicer input \cite{Cioffi95}.

Decision-feedback equalization (DFE) (see Figure~\ref{fig:DFE}) is based on using previously detected symbols in order to cancel the induced ISI from the symbol entering the slicer. In this approach, if all previous data symbols are detected without error, then postcursor ISI can be removed.
Specifically, the output of the FFE in Figure~\ref{fig:DFE} is given by
\begin{align}
y'_k &=x_k*g_k+z_k\nonumber \\
&=x_k+\sum_{m=-\infty}^{-1}x_{k-m} g_{m}+\sum_{m=1}^{\infty}x_{k-m} g_{m}+z_k\nonumber\\
&=x_k+\text{ISI}_k^{\text{pre}}+\text{ISI}_k^{\text{post}}+z_k,\nonumber
\end{align}
where $G(D)$ is the equivalent channel after the operation of the FFE.
The DFE then subtracts $\hat{\text{ISI}}_k^{\text{post}}=\sum_{m=1}^{\infty}\hat{x}_{k-m} g_{m}$ from $y'_k$, where $\hat{x}_k$ are decisions on past transmitted symbols, giving rise (assuming correct past decisions) to the equivalent channel
\begin{align}
y''_k=x_k+\sum_{m=-1}^{-\infty}x_{k-m} g_{m}+z_k.\nonumber
\end{align}
MMSE-DFE is optimal, when using the optimal FFE, in the sense that the SINR at the slicer (assuming correct past decisions) is equal to the SNR of an AWGN channel having the same capacity as that of the ISI channel \cite{Cioffi95}. Combining DFE with coding, however,  is a non-trivial task.
Since a decision on the value of the last symbol $\hat{x}_k$ must enter the feedback loop at every time instance, there is an intrinsic tension with the latency required for channel coding.  Many approaches have been suggested in order to overcome this obstacle (see for example \cite{amrani,wornellIterative,DFEcoded,DFEcoded2,DFEcoded3}), but to the best of our knowledge none of them allow to exchange the order of decoding and ISI removal which is the aim of the present work. Doing so, {\em directly} addresses the basic problem of ensuring that reliable decisions enter the DFE loop.

\section{Combining Cyclic Codes With IF Equalization}
\label{sec:2}

We begin this section by recalling IF equalization in the context of ISI channels. For a more general description, see \cite{IFmimo}.
For purposes of simplicity of exposition, we limit ourselves in the sequel to real-valued channels (and transmission of real symbols). The extension of the proposed scheme to complex transmission is straightforward, and, unless stated otherwise, all results derived in this work hold for the complex case as well. Further, we first describe the ZF-IF approach which is suitable for the high SNR regime, and relegate the extension to MMSE-IF (which is suitable for any SNR) to the Appendix.

An IF equalizer (depicted in Figure~\ref{fig:if}), rather than attempting to cancel the ISI (as in linear ZF as well as ZF decision feedback), ensures that the ISI is restricted to take only integer values. Assuming the data symbols are taken from a constellation of size $q$ consisting of the integers $\left\{0,1,\ldots,q-1\right\}$ (i.e., a PAM constellation), which we may identify as the ring $\mathbb{Z}_q$, the ISI will take only integer values provided that the channel impulse response is equalized to an impulse response ${I(D)}$, such that all the coefficients of ${I(D)}$ are integers. We restrict attention to
responses $I(D)$ of finite length and further assume that $I(D)$ is a monic\footnote{There is some loss of generality in this assumption. However, the optimal choice for $I(D)$ (as we discuss in Section~\ref{sec:3}) is almost always monic. When this isn't the case, the proposed approach is not very effective. The restriction that $I(D)$ be monic can be removed with some modifications to the scheme.}
 polynomial (with integer coefficients), i.e.,
$$I(D)=\sum_{k=0}^{n-1} i_k D^k,$$
where $i_0=1$. We denote the vector of coefficients of  $I(D)$ by,
$$\mathbf{i}=\left[1 \ i_1 \ \ldots \ i_{n-1}\right].$$
Note that taking ${I(D)}=1$ is a special case in which the IF equalizer reduces to a ZF-LE one. In most cases, choosing ${I(D)}$ otherwise results in smaller noise enhancement. The criterion for determining the feed-forward filter for the IF equalizer will be given in the next section.
We further note that IF equalization is closely related to partial response  signaling (see for example \cite{KP75}),  and was previously suggested in \cite{Fischer06}
where a suboptimal algorithm for finding a good integer channel $I(D)$ was also suggested. Another work of similar spirit is \cite{Zehavi} where it is shown that in some cases the complexity of the maximum likelihood decoder that is matched both to the channel and the code can be reduced if the channel is integer-valued.

The key feature of IF equalization exploited in this paper is that, as we observe below,
if the transmitted data is taken from a {\em cyclic} code, then the output of the equalized channel
is also a member of the codebook.

Assume some monic integer-valued filter
was chosen by the receiver (the considerations for choosing ${I(D)}$  are discussed in Section~\ref{sec:3}).
The FFE applied as the front-end filter is therefore,
\begin{equation}
A_{\text{ZF-IF}}(D)=\frac{I(D)}{H(D)}.
\label{eq:if_FE}
\end{equation}
The output of the equalizer is
\begin{eqnarray*}
Y'(D) & = &  X(D)I(D)+Z(D),\nonumber
\end{eqnarray*}
where $Z(D)=N(D) I(D)/H(D)$ is colored Gaussian noise.
We further define the SNR to be,
\begin{equation}
\snr_{\text{\rm ZF-IF-DFE}}=
\frac{\sigma_x^2}{\sigma_z^2}=
\frac{\sigma_x^2}{\frac{1}{2\pi}\int_{-\pi}^{\pi}\frac{|{I({e^{{jw}}})}|^{2}} {|{H({e^{{jw}}})}|^{2}}d \omega}.\nonumber
\end{equation}
The impulse response $I(D)$ should be chosen so as to maximize $\snr_{\text{\rm ZF-IF-DFE}}$. The definition of SNR (i.e, which does not include the residual ISI as noise) is justified when using a cyclic code as described next.
\vspace{2mm}
\begin{definition}
A linear block code $\mathcal{C}$ of length $N$ over $\mathbb{Z}_q$ is called cyclic, if for every codeword $\bx \in \mathcal{C}$, all cyclic shifts of $\bx$ are also codewords in $\mathcal{C}$.
\label{cyclicCode}
\end{definition}
\vspace{2mm}
We denote cyclic convolution w.r.t. block length $N$ by $\otimes$. The following proposition is an immediate consequence of Definition \ref{cyclicCode}.
\vspace{2mm}
\begin{proposition}
Let $\mathcal{C}$ be a cyclic code of length $N$ over $\mathbb{Z}_q$.
Then for any vector $\mathbf{i}$ of length $N$ with integer entries,
\[\mathcal{C} \otimes \mathbf{i} \subseteq \mathcal{C}. \]
That is, $\mathcal{C}$ is closed under integer-valued cyclic convolution over $\mathbb{Z}_{q}$.
\label{lemma_closed}
\end{proposition}
\vspace{2mm}
We conclude that if the linear convolution $x_k \ast i_k$ over $\mathbb{R}$ were a cyclic convolution $x_k \otimes i_k$ with operations over
 $\mathbb{Z}_q$,
then we would be able to decode $\bx'=\bx \otimes \mathbf{i}$ directly from the output $\mathbf{y'}$, and then reconstruct $\bx$ from it.\footnote{
It is important to note that in order to reconstruct $\mathbf{x}$ from $\bx \otimes {\bf i}\bmod {q}$, all elements of $\bx$ must be in $\mathbb{Z}_{q}$. Nevertheless, the elements of $\mathbf{i}$ need only be integers but need not necessarily be restricted to $\mathbb{Z}_{q}$.}

The transformation to cyclic convolution
may be accomplished in much the same way as in DMT/OFDM transmission.
In order to do so, we use a systematic cyclic encoder at the transmitter such that the last $K$ symbols of a ${[N,K]}$ (cyclic) block code are the data symbols and the first $N-K$ symbols are redundancy symbols. We fix the last $n-1$ information symbols to be zeros, where $n$ is the length of $I(D)$. Consequently, {\em all} codewords end with $n-1$ zeros. The effective rate is reduced to $(K-n+1)/N$.

Due to the zero padding, if we take only the first $N$ samples of the output of the FFE,
reduced modulo $q$, we get,
\begin{align}
y_k' \modulo q = \left[(\mathbf{x}\otimes\mathbf{i})_k + z_k \right] \modulo q, \ \ \ \mbox{for $k=1,\ldots,N$}.
\end{align}

\begin{figure}[htb]
      \begin{center}
                       \includegraphics[width=0.9\columnwidth]{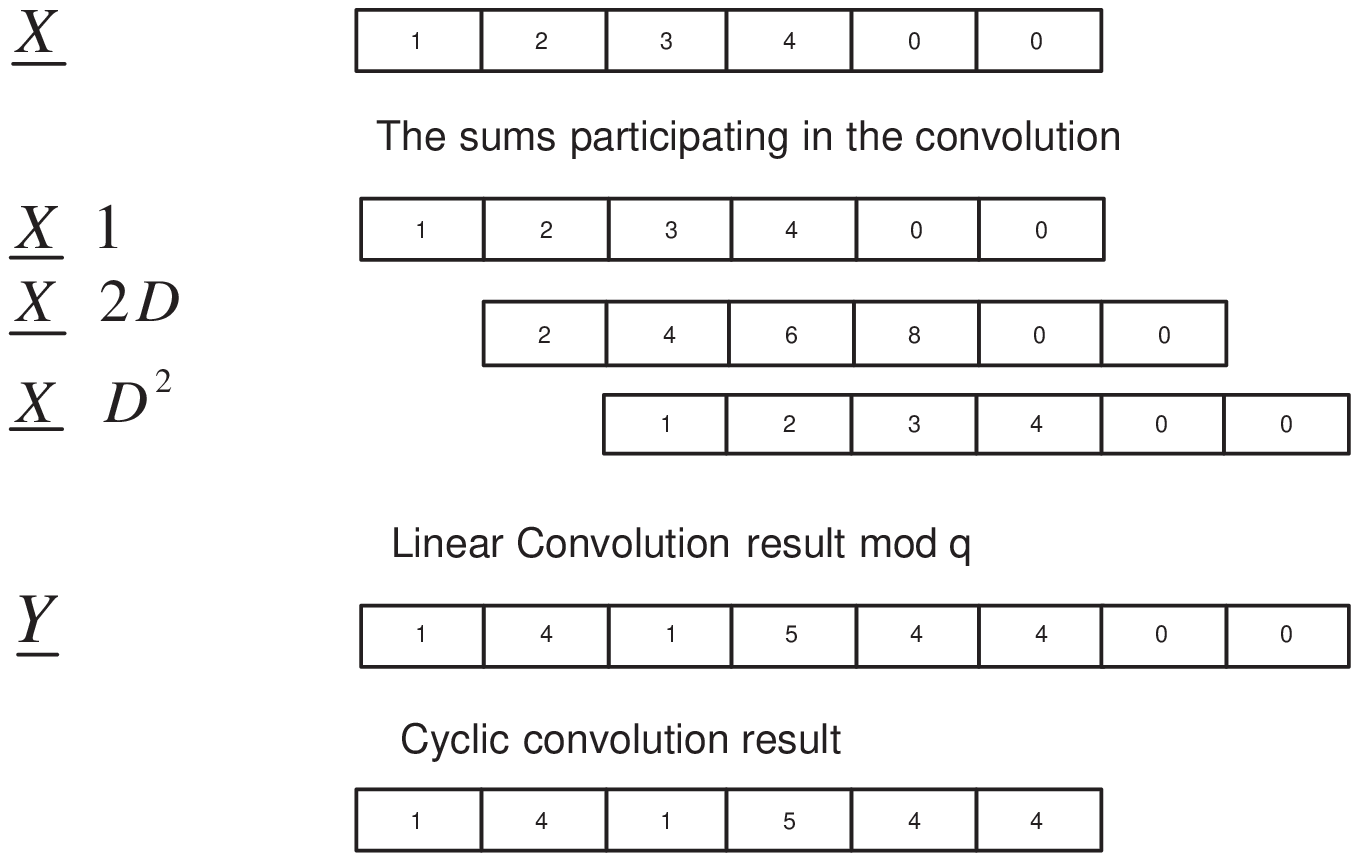}
      \end{center}
\caption{An Illustration of transforming a linear convolution ${x_k}\ast{i_k}$  over $\mathbb{R}$ into a cyclic convolution ${x_k}\otimes{i_k}$ over $\mathbb{Z}_{7}$. In this example ${I(D)}=1+2{D}+{D^{2}}$.}
  \label{zerosPaddingIlustration}
\end{figure}

This process is illustrated in Figure~\ref{zerosPaddingIlustration}.
The output of the FFE is processed in blocks of length $N$ (the blocks play no role in the operation of the FFE itself).
The zero padding, in addition to initializing the DFE with zeros, as described below, ensures that when a new data block is transmitted, the ``channel's memory'' is empty, i.e., there is no ISI between consecutive blocks.

Since the result of the cyclic convolution is itself a codeword, we may apply the channel decoder directly to it. Assuming correct decoding, we next reconstruct ${x_k}$ from the decoded result of the cyclic convolution. This can be done in a recursive way by applying a DFE as shown in Figure~\ref{fig:if}.\footnote{All operations of the DFE are carried over $\mathbb{Z}_q$ It is always possible to reconstruct $x_k$ regardless of $q$ (i.e. whether it is prime or not) and the choice of the integer-valued filter since $I(D)$ is monic.}
Since we know the last $n-1$ values of $x_k$ (i.e., they are all zero), we are able to recover $x_k$ for every $k=1,...,N-(n-1)$.

The reconstruction of $x_k$ from $x_k\otimes i_k$ is performed using a DFE rather than a linear equalizer, as the integer filter $I(D)$ often has zeros on the unit circle, and hence it is not invertible.

\vspace{2mm}

We end this section by remarking that the recently proposed ``Signal Codes''~\cite{SignalCodes}, though not cyclic, are also suitable for IF equalization. Signal codes are a special family of lattice codes which are generated by convolving sequences of data symbols taken from a PAM/QAM constellation with a monic filter $F(D)$. For data sequences of length $N$ and an encoding filter $F(D)$ of length $M$, the generating matrix of the lattice code is an $(N+M-1)\times N$ Toeplitz matrix.
For a data sequence $B(D)$ the corresponding codeword is\footnote{In fact, some additional ``shaping'' steps must be taken in order to assure that the power constraint is preserved. The full details can be found in~\cite{SignalCodes}.} $X(D)=B(D)F(D)$. When IF equalization is used in conjunction with signal codes, the output of the equivalent channel after the FFE is (neglecting the additive noise) $X'(D)=B(D)I(D)F(D)$. Let $B'(D)=B(D)I(D)$, and note that $B'(D)$ is a sequence of $(N+n-1)$ PAM/QAM symbols as $I(D)$ is an integer valued filter of length $n$. It follows that $X'(D)$ is a codeword from a signal code generated by $F(D)$ with a generating Toeplitz matrix of dimensions $(N+M+n-2)\times (N+n-1)$. For large enough $N$, this codebook has similar performance to those of the original codebook. Therefore $B'(D)$ can be decoded and then $B(D)$ can be reconstructed from it.
Note that for signal codes there is no need to transform the linear convolution performed by the channel into a cyclic one, as done when cyclic codes are used.
In~\cite{SignalCodes} it is demonstrated that signal codes can operate reasonably close to capacity with an acceptable complexity, and may therefore be attractive for IF equalization at high SNR.

\section{Criteria for Choosing The Filter ${I(D)}$}
\label{sec:3}

We wish to find an integer-valued filter $I(D)$ of length $n$ such that the noise enhancement experienced by the ZF-IF equalizer for a given channel ${H(D)}$ is minimized. The noise variance at the output of the IF front-end filter (\ref{eq:if_FE}) is,
\begin{eqnarray}
{\sigma_{\text{\rm ZF-IF-DFE}}^{2}}={\frac{1}{2\pi}}\int_{-\pi}^{\pi}\frac{|{I({e^{{j\omega}}})}|^{2}} {|{H({e^{{j\omega}}})}|^{2}}d\omega.            \label{eq11}
\end{eqnarray}
Denote,
\begin{equation}
K(D)=\frac{1}{H(D)H^*(D^{-*})}.
\label{eqKD}
\end{equation}
Thus, $k_m$ is an autocorrelation sequence (and in particular $k_{-m}=k_m$).

Using straightforward algebra, (\ref{eq11}) may be written as a quadratic form,
\begin{align}
{\sigma_{\text{\rm ZF-IF-DFE}}^{2}}&={\mathbf{i}}{{\tilde{\text{K}}}_n}{\mathbf{i^{T}}},\nonumber
\end{align}
where $\tilde{\text{K}}_n$ is the positive semi-definite Toeplitz matrix,
\begin{align}
{\tilde{\text{K}}}_n=
\left[\begin{array}{ccccc}
k_{0} & k_{-1} & k_{-2} \ldots & k_{-(n-1)}\\
k_{1} & k_{0} & k_{-1} \ldots & k_{-(n-2)}\\
\vdots & \vdots & \vdots \ddots\\
k_{n-1} & k_{n-2} & k_{n-3} \ldots & k_{0}
\end{array}\right].\nonumber
\end{align}
Let $\text{F}_n$ be any matrix satisfying,
\begin{eqnarray}
\tilde{\text{K}}_n={{\text{F}}_n}^T {\text{F}}_n.\nonumber
\end{eqnarray}
We therefore have,
\begin{eqnarray}
{\sigma_{\rm ZF-IF-DFE}^{2}}={\mathbf{i}}{\mathbf{\tilde{\text{K}}}}_n{\mathbf{i^{T}}}=
{\mathbf{i}}{\text{F}_n}^{T}{\text{F}_n}{\mathbf{i^{T}}}=
\|{\text{F}_n}{\mathbf{i^{T}}}\| ^{2}.\
            \label{eq18}
\end{eqnarray}
Equation~(\ref{eq18}) implies that finding the optimal (ZF) integer-valued filter $I(D)$ is equivalent to finding the shortest vector in the lattice $\Lambda({\text{F}}_n)$, which is composed of all integral combinations of the columns of ${\text{F}_n}$, i.e.,
\begin{eqnarray}
\Lambda({\text{F}}_n)=\{\mathbf{\lambda}={\text{F}_n}\mathbf{i} \ | \ \mathbf{i}\in \mathbb{Z}^{n}\}.
\end{eqnarray}
Finding the shortest lattice vector is known to be NP hard, but fortunately efficient suboptimal algorithms for finding a short lattice basis are known. An important representative of this class of algorithms is the celebrated LLL algorithm \cite{LLL}, which has polynomial complexity and usually gives adequate results in practice. In order to find a ``good" integer-valued filter $I(D)$, we may therefore apply the LLL algorithm on ${\text{F}_n}$. The algorithm's result is a new basis (of short vectors) for ${\text{F}_n}$. We then need to find the shortest vector $\mathbf{v}$ in this basis and choose $\mathbf{i}={\text{F}_n}^{-1}\mathbf{v}$.
In the next section we derive an upper bound on the induced noise enhancement when the true (optimal) shortest vector of the lattice is used, which serves as a useful benchmark.

\section{An Upper Bound On The Noise Enhancement of optimal ZF-IF Equalization}
\label{sec:4}
We now upper bound the noise enhancement induced by optimal ZF-IF equalization.
Throughout this Section we assume that the channel has a finite-length impulse response of length $p+1$, i.e.,
\[H(D)=\sum_{k=0}^p h_k D^k.\]
The length of the channel will play an important role in the derived bound.

We proceed to upper bound ${\sigma_{\text{\rm ZF-IF-DFE}}^{2}}$ using known results from the theory of lattices, and the theory of Toeplitz matrices.
\vspace{2mm}
\begin{lemma}{(Minkowski)}
Let $\lambda_{1}\left(\Lambda\left({\text{F}_n}\right)\right)$ denote the shortest vector in the $n$-dimensional lattice $\Lambda\left({\text{F}_n}\right)$. If ${\text{F}_n}$ is of rank $n$, then,
\begin{eqnarray}
\lambda_{1}\left(\Lambda\left({\text{F}_n}\right)\right)\leq \left(\frac{2^{n}}{\beta_{n}}\right)^{\frac{1}{n}}\left|\det\left({\text{F}_n}\right)\right|^{\frac{1}{n}},\nonumber
\end{eqnarray}
where $\beta_{n}$ is the constant of proportionality of the volume of an $n$-dimensional sphere, i.e., the volume of an $n$-dimensional sphere with radius $R$ is given by $V_{n}(R)=\beta_{n}R^{n}$.
\label{MinLem}
\end{lemma}
\vspace{2mm}
\begin{proof}
See, e.g., \cite{latticeBook}.
\end{proof}
The constant $\beta_{n}$ can be bounded by (see, e.g., \cite{ConwaySloane}),
\begin{eqnarray*}
\beta_n>\left(\frac{2\pi e}{n}\right)^{n/2}\frac{1}{\sqrt{1.4\pi n}}.
\end{eqnarray*}
Recalling that $\tilde{\text{K}}_n={\text{F}_n}^T {\text{F}_n}$, we therefore have,
\begin{eqnarray}
\lambda^2_{1}\left(\Lambda\left({\text{F}_n}\right)\right)< \eta(n) \cdot \left[\det(\tilde{\text{K}}_n)\right]^{\frac{1}{n}},
 \label{eq19}
\end{eqnarray}
where
\begin{equation}
\eta(n) = {\frac{2n}{\pi e}} \cdot \left(1.4\pi n\right)^{\frac{1}{n}}.
\label{eq:eta}
\end{equation}
Therefore, in order to obtain  a closed-form bound on $\lambda^2_{1}\left(\Lambda\left({\text{F}}_n\right)\right)$, it suffices to evaluate $\det\left(\tilde{\text{K}}_n\right)$.
To that end, define,
\begin{equation}
\alpha_H=\frac{|z_{0}z_{1} \ldots z_{p-1}|^{2p}}{\prod_{{\mu ,\nu}}|z^{*}_{\mu}z_{\nu}-1|},
\label{eqalpha}
\end{equation}
where $z_{0},z_{1},\ldots,z_{p-1}$ are the maximum-phase zeros of $H(D)H^*(D^{-*})$ (i.e., the zeros outside the unit circle).

We are now ready to present Theorem~\ref{Bound} which is the main result of this section.
The theorem makes use of a result from the theory of Toeplitz matrices which we state first, and the
proof of which can be found in~\cite[Chapter 5]{ToeplitzBook}.
\vspace{2mm}
\begin{lemma}{}
Let $H(D)=\sum_{k=0}^p h_k D^k$ be a polynomial of degree $p$.
Further, let $K(D)$ be as defined in (\ref{eqKD}).
Then for every $n\geq p+1$,
\begin{eqnarray*}
\left[\det(\tilde{\text{K}}_{n})\right]^{\frac{1}{n}}=(\alpha_H)^\frac{1}{n} \cdot \exp\bigg[{\frac{1}{2\pi}\int_{-\pi}^{\pi}\log K(e^{j \omega}) d\omega}\bigg],
\end{eqnarray*}
where $\alpha_H$ is defined in (\ref{eqalpha}). 
\label{ToepLem}
\end{lemma}
\vspace{2mm}

\vspace{2mm}
\begin{theorem}
Assume that the channel has a finite-length impulse response $H(D)=\sum_{k=0}^p h_k D^k$. For an optimal choice of the integer filter $I(D)$, the noise power at the output of the filter ${I(D)}/{H(D)}$ is bounded by\footnote{For (complex) transmission over a complex channel, (\ref{eq:eta}) changes to:
\begin{align*}
\eta(n) = \frac{4n}{\pi e} \cdot \left(2.8\pi n\right)^{\frac{1}{2n}}.
\end{align*}
}
\begin{align}
\sigma^{2}_{\text{\rm ZF-IF-DFE}}\leq  \sigma^{2}_{\rm ZF-DFE} \cdot \min_{n\geq p+1}\bigg[\eta(n)  \cdot (\alpha_H)^{\frac{1}{n}}\bigg ] ,
 \label{eq20}
\end{align}
where $\eta(n)$ is defined in (\ref{eq:eta}) and $\alpha_H$ is defined in (\ref{eqalpha}).
\label{Bound}
\end{theorem}
\vspace{2mm}
\begin{proof}
Lemma~\ref{ToepLem} may be rewritten as 
\begin{align}
\left[\det({{\tilde{\text{K}}}}_n)\right]^{\frac{1}{n}}=&\left(\alpha_H\right)^{\frac{1}{n}}\cdot
\sigma^{2}_{\rm ZF-DFE},
\label{Det}
\end{align}
where
\begin{align}
\sigma^{2}_{\rm ZF-DFE}=\exp\bigg[-{\frac{1}{2\pi}\int_{-\pi}^{\pi}\log{|{H({e^{{j\omega}}})}|^{2}}d\omega}\bigg]
\label{NHdfe}
\end{align}
is the variance of the AWGN after the optimal FFE is applied in ZF-DFE (see, e.g., \cite{Cioffi95}).
Substituting (\ref{Det}) into (\ref{eq19}) yields:
\begin{align*}
\lambda_{1}(\Lambda\left(\text{F}_n\right))^{2}\leq & \eta(n) \cdot \left(\alpha_H\right)^{\frac{1}{n}}
\cdot \sigma^{2}_{\rm ZF-DFE}.\nonumber
\end{align*}
Since $\sigma^{2}_{\text{\rm ZF-IF-DFE}}=\lambda^2_{1}\left(\Lambda\left({\text{F}}_n\right)\right)$, the theorem is proved.
\end{proof}

The additional noise enhancement caused by the ZF-IF equalizer w.r.t. an optimal ZF-DFE as bounded in (\ref{eq20}) consists of two factors: $\left(\alpha_H\right)^{\frac{1}{n}}$ and $\eta(n)$. The factor $\left(\alpha_H\right)^{\frac{1}{n}}$ is greater (or equal) to $1$ and tends to $1$ as $n \rightarrow \infty$. Thus, allowing for a long filter $I(D)$ mitigates the effect of this factor. On the other hand, the factor $\eta(n)$ is (approximately) linearly increasing with $n$. The minimization in (\ref{eq20}) strikes a balance (i.e., searches for the optimal tradeoff) between
these two factors.

The tightness of the bound depends on the tightness of the Minkowski bound of Lemma~\ref{MinLem}. It is known that there exist lattices that satisfy this bound with equality, but for ``most" lattices the shortest lattice vector is much shorter than what this bound predicts. We further note that the family of lattices considered in this paper is not general, as $\text{F}_n^T \text{F}_n$ is a Toeplitz matrix, and therefore the Minkowski bound may never be tight, as we further discuss in the next section.

\section{Performance of ZF-IF Equalization}
\label{sec:performance}

The capacity of the Gaussian ISI channel~(\ref{eq2b}) at high SNR is given by (see e.g.~\cite{Cioffi95})
\begin{align}
C=\frac{1}{2}\log_2\left(\frac{\text{SNR}}{\sigma^2_{\text{ZF-DFE}}}\left(1+o(1)\right)\right),
\label{ISIcapacity}
\end{align}
where $\sigma^2_{\text{ZF-DFE}}$ was defined in~(\ref{NHdfe}), and where $o(1)\rightarrow 0$ as $\text{SNR}\rightarrow\infty$. Note that~(\ref{ISIcapacity}) is valid only for channels for which $\sigma^2_{\text{ZF-DFE}}$ is finite, i.e. for channels satisfying the Paley-Wiener condition (see~\cite{Lee&Mess}).

In this section we analyze the total gap-to-capacity of the ZF-IF equalization scheme at high SNR, i.e. the gap between the performance obtained using ZF-IF and the optimal performance~(\ref{ISIcapacity}).

The ZF-IF scheme, as described in Section~\ref{sec:2}, transforms the original Gaussian ISI channel into an equivalent modulo-additive channel
\begin{align}
Y'(D)&=\left[X(D)I(D)+Z(D)\right]\bmod q\nonumber\\
&=\left[X'(D)+Z(D)\right]\bmod q,\nonumber
\end{align}
where $Z(D)$ is filtered Gaussian noise. As we recall, correct decoding of $X'(D)$ ensures correct reconstruction of $X(D)$.
In practice, the transmitted signal $x_k$ is subject to the power constraint
\begin{align}
E\left[x_k^2\right]\leq\text{SNR},\nonumber
\end{align}
and hence, instead of mapping the codewords from the cyclic code $\mathcal{C}$ to the constellation $\mathbb{Z}_q$, we map them to the constellation
\begin{align}
{c\sqrt{\text{SNR}}}\left\{\frac{-(q-1)}{2q},\frac{-(q-3)}{2q},\ldots,\frac{q-1}{2q}\right\}.
\label{constellation}
\end{align}
The constant $c$ is chosen such that the power constraint is satisfied with equality for a uniform distribution over the constellation. It can be easily verified that $c>\sqrt{12}$ for any value of $q$ and approaches $\sqrt{12}$ for large values of $q$.\footnote{The codebook $\mathcal{C}$ is invariant to the convolution with $I(D)$ as defined in Section~\ref{sec:2}, and hence $X'(D)$ is uniformly distributed over the constellation~(\ref{constellation}), as is $X(D)$.}

Let $\Delta=c\sqrt{\text{SNR}}$. Using the constellation~(\ref{constellation}), the equivalent modulo-additive channel after the FFE can be rewritten as
\begin{align}
Y'(D)=\left[X'(D)+Z(D)\right]\bmod [-\Delta/2,\Delta/2).
\label{modAdditiveChannel}
\end{align}
In order to analyze the performance limits of the equivalent channel, we lower bound the mutual information $I(X';Y')$ between its input and output. The mutual information corresponding to a certain distribution on the channel's input gives the highest possible rate for reliable communication, when a random channel code which is drawn according to that distribution is used~\cite{Cover}. However, the equivalent channel~(\ref{modAdditiveChannel}) is attained using a linear cyclic code (as opposed to a random code), and thus the mutual information may never be achieved. Nevertheless, the mutual information is a useful metric for the  performance of ZF-IF if we account for the possible loss of rate due to the use of a linear cyclic code.

The channel~(\ref{modAdditiveChannel}) is a modulo-additive \emph{colored} Gaussian noise channel, where the variance of the noise is $\sigma^{2}_{\text{\rm ZF-IF-DFE}}$. The mutual information between the input and output of such a channel is maximized by a uniform memoryless distribution over the modulo interval. Moreover, for such an input distribution, the mutual information is lower bounded by that of a modulo-additive channel with \emph{white} Gaussian noise having the same variance, see~\cite{DPC}. Let $\tilde{Z}$ be a Gaussian random variable with zero mean and variance $\sigma^{2}_{\text{\rm ZF-IF-DFE}}$. If $X'$ has a memoryless uniform distribution over $[-\Delta/2,\Delta/2)$, which corresponds to taking $q$ to infinity, $Y'$ has a memoryless uniform distribution over the interval as well, and we have
\begin{align}
I(X';Y')&\geq\log_2(\Delta)-h(\tilde{Z} \bmod [-\Delta/2,\Delta/2))\nonumber\\
&\geq \log_2(\Delta)-h(\tilde{Z})\label{moduloReduces}\\
&=\frac{1}{2}\log_2(\Delta^2)-\frac{1}{2}\log_2(2\pi e\cdot \sigma^{2}_{\text{\rm ZF-IF-DFE}})\nonumber\\
&=\frac{1}{2}\log_2\left(\frac{\text{SNR}}{\sigma^{2}_{\text{\rm ZF-IF-DFE}}}\cdot\frac{12}{2\pi e}\right),\nonumber
\end{align}
where~(\ref{moduloReduces}) follows from the fact that modulo reduction can only decrease differential entropy.
Let us express $\sigma^{2}_{\text{\rm ZF-IF-DFE}}$ as a product of two factors
\begin{align}
\sigma^{2}_{\text{\rm ZF-IF-DFE}}=\gamma_{I,H}\cdot\sigma^2_{\text{ZF-DFE}},\nonumber
\end{align}
where $\gamma_{I,H}$ is the additional noise enhancement caused by the FFE in ZF-IF w.r.t. that caused by the FFE of ZF-DFE. With this notation we have
\begin{align}
I(X';Y')&\geq \frac{1}{2}\log_2\left(\frac{\text{SNR}}{\sigma^{2}_{\text{\rm ZF-DFE}}}\cdot\frac{1}{\gamma_{I,H}}\cdot\frac{12}{2\pi e}\right)\nonumber\\
&=\frac{1}{2}\log_2\left(\frac{\text{SNR}/\Gamma_{I,H}}{\sigma^{2}_{\text{\rm ZF-DFE}}}\right),\nonumber
\end{align}
where \[\Gamma_{I,H}=\frac{2\pi e}{12}\cdot{\gamma_{I,H}}\] is the SNR loss due to IF-ZF equalization. It follows that the gap-to-capacity in dB is given by\footnote{This is the gap-to-capacity in the case of (complex) transmission over complex channels as well.}
\begin{align}
10\log_{10}(\Gamma_{I,H})=10\log_{10}\left(\frac{2\pi e}{12}\right)+10\log_{10}(\gamma_{I,H}).
\label{GapToCapac}
\end{align}
The first term on the r.h.s. of~(\ref{GapToCapac}) is the well-known high-SNR shaping gain, which equals $1.53$dB. The reason for this loss is the one-dimensional modulo operation we use at the receiver.
The modulo operation allows the decoder to use the original codebook $\mathcal{C}$ for decoding $\mathbf{x'}$ which preserves the original decoding complexity of the codebook. However, decoding the equivalent channel's output after the modulo reduction is equivalent to searching for the point that was most likely transmitted over the \emph{infinite} lattice $\mathcal{C}+\Delta\mathbb{Z}^N$.
This is strictly suboptimal since no more than $2^{NR}$ points of the infinite lattice, which correspond to $\mathcal{C}\otimes\mathbf{i}$ (with operations carried over the \emph{reals}), are valid, and hence better performance can be achieved by searching only over these valid points at the decoder.

While the modulo loss amounts to $1.53$dB at high-SNR, at low SNR the loss may be significantly larger (see, e.g., \cite{DPC}), which makes IF equalization less attractive in that regime. This loss can be mitigated by incorporating shaping at the transmitter, and/or avoiding the modulo reduction at the receiver. Implementing such modifications at the transmitter and receiver without significantly increasing the computational complexity is an interesting avenue for future research.

\vspace{2mm}

The second loss in~(\ref{GapToCapac}) is related to the additional noise enhancement caused by the FFE. It follows from Theorem~\ref{Bound}, that for the optimal choice of the integer-valued filter, $\gamma_{H}$ is upper bounded by
\begin{align}
\gamma_{H}=\min_{I(D)}\gamma_{I,H}\leq\min_{n\geq p+1}\bigg[\eta(n)  \cdot (\alpha_H)^{\frac{1}{n}}\bigg ].
\label{gammaBound}
\end{align}
In order to gain more insight regarding the term $\gamma_{I,H}$ we illustrate the effect of the noise enhancement through the following examples of ISI channels.

\begin{figure}[htb]
 \includegraphics[width=1\columnwidth]{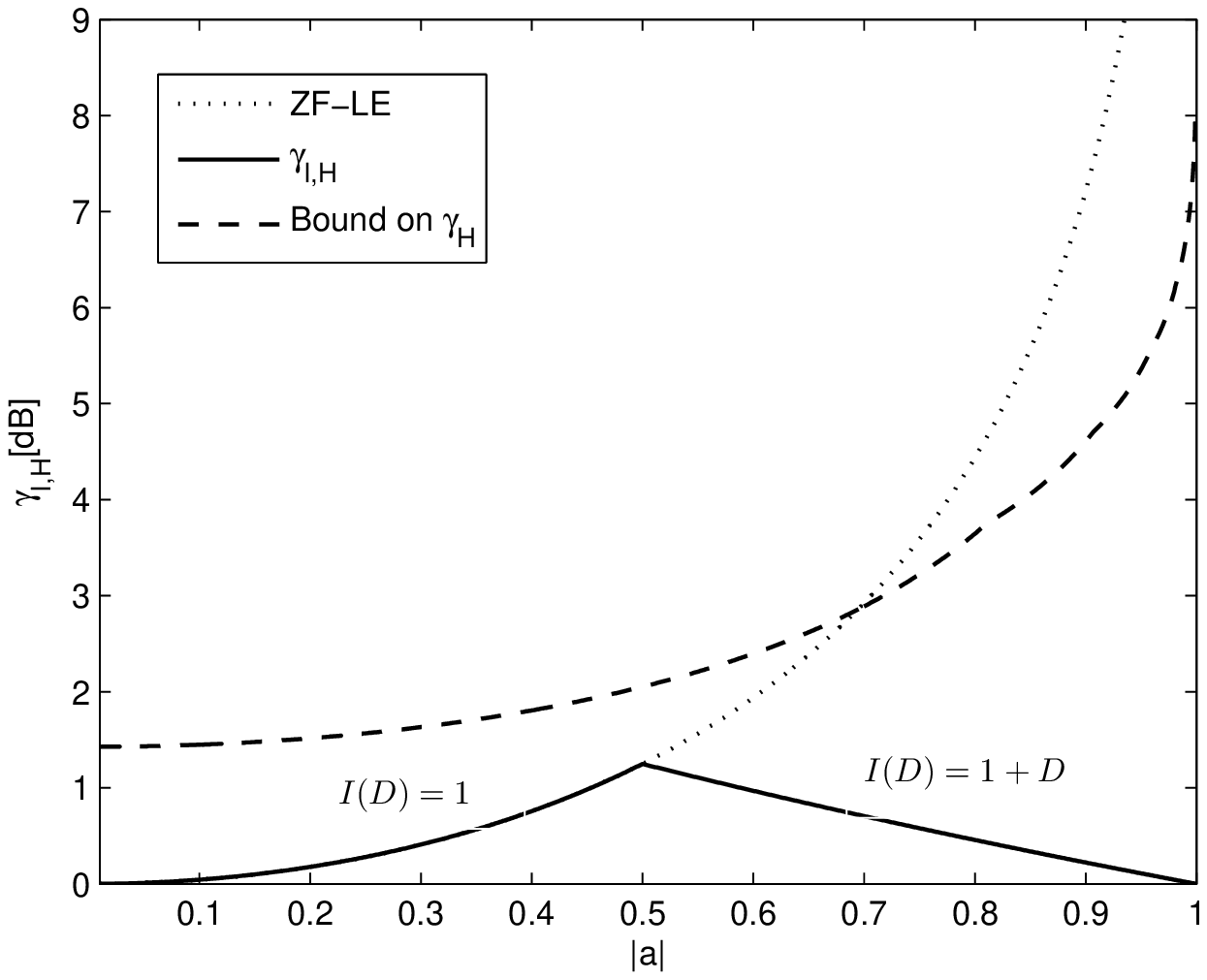}
\caption{$\gamma_{I,H}$ for the real two-tap channel $H(D)=1+aD^p$. The solid line shows $\gamma_{I,H}$ for the optimal choice of $I(D)$ for \emph{any} $p$, the dashed line shows the bound~(\ref{gammaBound}) for $p$=1 and the dotted line shows the noise enhancement induced by ZF-LE}
 \label{MinkowskiBound}
\end{figure}

{\bf Example $1$ - Two-tap real ``RAKE" channel:}
An interesting example is a {\emph{real}} channel with only two non-zero taps (which may be arbitrarily far apart), namely $H(D)=1+aD^p$. Without loss of generality, we can assume that $|a|\leq 1$, as otherwise we can transform the channel into the channel $\tilde{H}(D)=1+\frac{1}{a}D^p$ by using an all-pass filter. It is rather obvious that the optimal choice for $I(D)$ is either $I_0(D)=1$ or $I_1(D)=1+\frac{a}{|a|}D^p$. It can be shown by straightforward algebra that for the choice $I_0(D)$, the noise enhancement is given by $1/(1-a^2)$, and for the choice $I_1(D)$ the noise enhancement is given by $2/(1+|a|)$. It follows that for $|a|\leq 1/2$, $I_0(D)$ is better, while for $1/2<|a|\leq 1$, $I_1(D)$ is better, and the maximum noise enhancement (which occurs for $|a|=1/2$) is $4/3\approx 1.25$dB. Since for this channel $\sigma^{2}_{\text{\rm ZF-DFE}}=1$, $\gamma_{I,H}$ is equal to the noise enhancement. Figure~\ref{MinkowskiBound} depicts $\gamma_{I,H}$ for the real two-tap channel with an optimal choice of $I(D)$, along with the bound~(\ref{gammaBound}) for the same channel, and the noise enhancement caused by a ZF-LE. It is evident from the figure that for this channel the bound is not tight.

\begin{figure}[htb]
 \includegraphics[width=1\columnwidth]{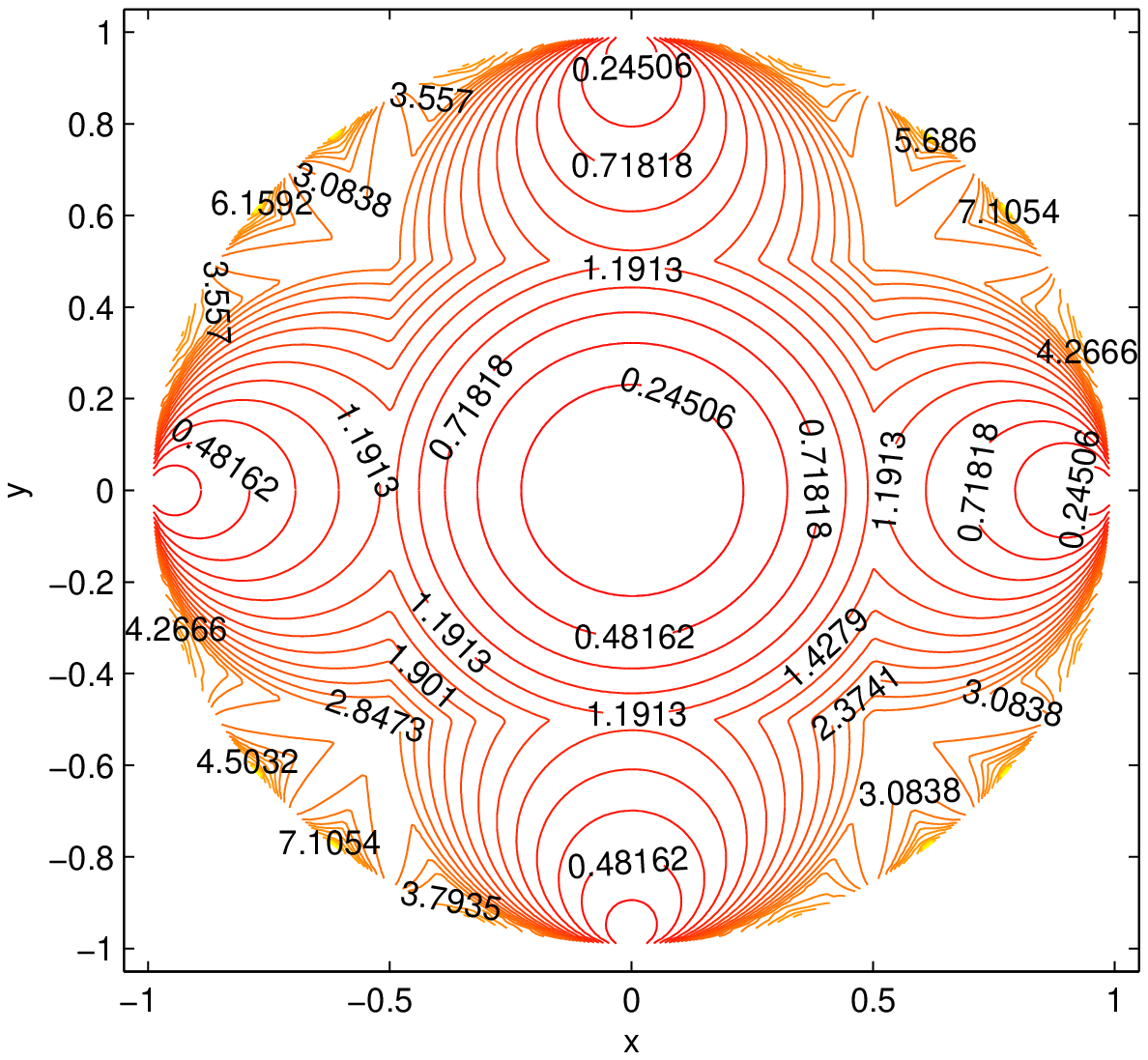}
\caption{$\gamma_{I,H}$ (in dB) for the complex two-tap channel $H(D)=1+(x+j\cdot y)D^p$, when the optimal integer-valued filter $I(D)$ is used.}
 \label{ComplexGamma}
\end{figure}

{\bf Example $2$ - Two-tap complex ``RAKE" channel:}
Consider the channel from the previous example $H(D)=1+a D^p$, where now $a=r e^{j\theta}$ is a complex number. We assume without loss of generality that $|a|\leq 1$. In contrast to the real-valued two-tap channel, in this case there is no clear choice of $I(D)$ for each value of $a$.
Nevertheless, similar to the real two-tap channel, $\gamma_{H}$ is independent of the delay $p$ between the first and the second tap (as is also the case for the ZF-LE). This follows since the noise enhancement caused by each of the filters $A_1(D)=I(D)/H(D)$ and $A_2(D)=I(D^p)/H(D^p)$ is the same. Therefore, if for the channel $H(D)=1+a D$ a certain choice of $I(D)$ results in noise enhancement of $\gamma_{I,H}$, for the channel $\tilde{H}(D)=1+a D^p$, the choice $\tilde{I}(D)=I(D^p)$ results in $\gamma_{I,H}$ as well.
While the performance of a ZF-LE is independent of the phase $\theta$, the noise enhancement caused by the ZF-IF equalizer is significantly influenced by $\theta$. It is interesting to note, however, that for the two-tap channel, the bound~(\ref{gammaBound}) (in its complex form) is independent of $\theta$. Figure~\ref{ComplexGamma} depicts $\gamma_{I,H}$ (in dB) for the two-tap channel with an optimal choice of $I(D)$ (which was found numerically), for values of $|a|\leq 0.99$.

\begin{figure}[htb]
 \includegraphics[width=1\columnwidth]{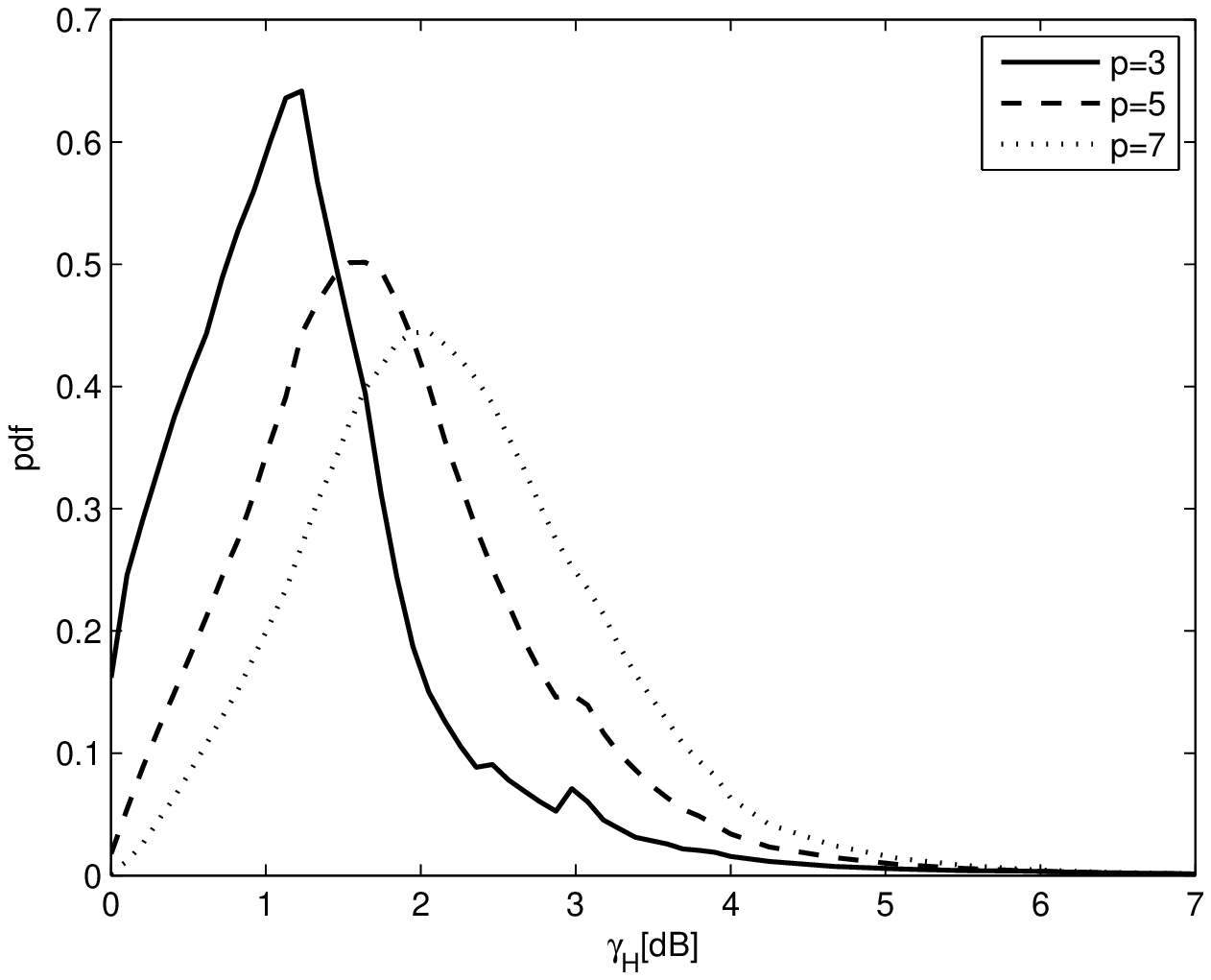}
\caption{The pdf of $\gamma_{I,H}$ (in dB) for the ISI channel $H(D)=\sum_{k=0}^p h_k D^p$ where $\left\{h_k\right\}_{k=0}^{p}$ are i.i.d. random Gaussian variables with zero mean and unit variance, when the optimal integer-valued filter $I(D)$ is used.}
 \label{randomTaps}
\end{figure}

{\bf Example $3$ - An ISI channel with random taps:}
In this example we consider the channel
\begin{align}
H(D)=\sum_{k=0}^p h_k D^p,\nonumber
\end{align}
where $\left\{h_k\right\}_{k=0}^{p}$ are real i.i.d. random Gaussian variables with zero mean and unit variance. We numerically evaluate the probability density function (pdf) of $\gamma_{I,H}$ in dB for the optimal choice of $I(D)$. Figure~\ref{randomTaps} depicts the results for $p=3$, $p=5$ and $p=7$. The results show that $\gamma_{H}$ becomes larger when the channel is longer, which is also the behavior of the bound~(\ref{gammaBound}).

\vspace{2mm}

A third loss incurred by ZF-IF equalization is related to the constraint that the channel code is cyclic. The gap-to-capacity of ZF-IF equalization at a certain block error probability, is the sum of the shaping loss, the additional noise enhancement $\gamma_{I,H}$ (in dB) and the gap-to-capacity (in dB) of the chosen cyclic code at the desired block error probability. To the best of the authors knowledge, it is still an open question whether cyclic codes can attain the capacity of the AWGN channel, let alone, that of the mod-$\Delta$ AWGN channel. Nevertheless, algebraic cyclic codes (such as BCH for example) have been extensively used in communication systems for decades due to the fair tradeoff they offer between performance and complexity. Modern (e.g., LDPC) cyclic codes are a subject of extensive research nowadays, and some families of such codes were reported to have very good performances over the AWGN channel\footnote{Note that in ZF-IF the additive noise after the FFE is not white, and thus it is important that the cyclic code should not be sensitive to the memory of the noise.}, see e.g.~\cite{Fossorier01,Idempotent}.

Most of the research effort on linear cyclic codes have been devoted to binary codes. In the high-SNR regime, which is the focus of this paper, the cardinality of the code must be greater than two, and a binary code does not suffice. In the next section we propose a practical coded modulation scheme for high transmission rates that utilizes binary cyclic codes and is suitable for cyclic-coded integer forcing equalization.

\section{Practical coding at high rates}
\label{sec:6a}

In order to achieve high transmission rates, using a binary channel codebook, we transmit a combination of coded and uncoded bits which are mapped to a $2^{M}$-PAM constellation points by {\em natural labeling}. Namely, we transmit,\footnote{Due to the power constraint, the transmitted signal, in effect, would be given by~(\ref{constellation}) with $q=2^M$.}
\begin{align}
\mathbf{x}=\mathbf{x}_c+\sum_{b=1}^{M-1} \mathbf{x}_{u_b}2^b,
\label{naturalLabeling}
\end{align}
where $\mathbf{x}_c$ is a codeword of length $N$ from the binary linear cyclic code $\mathcal{C}$, and $\mathbf{x}_{u_b} ,  \ b=1,\ldots,M-1$, are blocks of uncoded information bits having the same length ($N$) as that of $\mathbf{x}_c$.
Our goal is to detect $\left(\mathbf{x}\otimes\mathbf{i}\right)\modulo 2^{M}$ without error, and apply the DFE (with all operations carried over $\mathbb{Z}_{2^{M}}$) in order to reconstruct $\mathbf{x}$.\footnote{Note that in order to reconstruct $\mathbf{x}$ we need to know the value of its last $L$ samples. We therefore need to zero-pad the last $L$ uncoded bits as well as the last $L$ coded bits.}

We now show that natural labeling retains the closure property under integer-valued cyclic convolution.
\vspace{2mm}
\begin{lemma}
Let $\Lambda$ denote the set of vectors $\mathbf{x}$ given by (\ref{naturalLabeling}) such that $\mathbf{x}_c\in\mathcal{C}$. Specifically,
\begin{align}
\Lambda=\left\{\mathbf{x} \ | \ \mathbf{x}_c\in\mathcal{C},\mathbf{x}_{u_b}\in\mathbb{Z}_2^{N} ,  \ b=1,\ldots,M-1\right\}.\nonumber
\end{align}
If $\mathbf{x}\in\Lambda$, then $\left(\mathbf{x}\otimes\mathbf{i}\right)\modulo 2^M\in\Lambda$.
\end{lemma}
\vspace{2mm}
\begin{proof}
Using natural labeling, it is clear that the result of the cyclic convolution reduced modulo $2^{M}$ can be written as,
\begin{align}
\left(\mathbf{x}\otimes\mathbf{i}\right)\modulo 2^M=\mathbf{x}_{0}+\sum_{b=1}^{M-1} \mathbf{x}_{b}2^b\nonumber,
\end{align}
for some vectors $\mathbf{x}_{0},\mathbf{x}_1,\ldots,\mathbf{x}_{M-1}\in\mathbb{Z}_2^N$. Clearly if $\mathbf{x}_{0}\in\mathcal{C}$ then $\left(\mathbf{x}\otimes\mathbf{i}\right)\modulo 2^M\in\Lambda$. We have,
\begin{align}
\mathbf{x}_{0}&=\left(\left(\mathbf{x}\otimes\mathbf{i}\right)\modulo 2^M\right)\modulo 2\nonumber \\
&=\left(\mathbf{x}\otimes\mathbf{i}\right)\modulo 2\nonumber \\
&=\left(\left(\mathbf{x}_c+\sum_{b=1}^{M-1} \mathbf{x}_{u_b}2^b\right)\otimes\mathbf{i}\right)\modulo 2\nonumber\\
&=\left(\mathbf{x}_c\otimes\mathbf{i}+\sum_{b=1}^{M-1}2^b\left(\mathbf{x}_{u_b}\otimes\mathbf{i}\right)\right)\modulo 2 \nonumber\\
&\stackrel{(a)}{=}\left(\mathbf{x}_c\otimes\mathbf{i}\right)\modulo 2\stackrel{(b)}{\in}\mathcal{C},\nonumber
\end{align}
where $(a)$ holds since $\left(\mathbf{x}_{u_b}\otimes\mathbf{i}\right)\in\mathbb{Z}^N$, and $(b)$ holds because $\mathcal{C}$ is a binary linear cyclic codebook. Thus, $\mathbf{x}_0$ is indeed a codeword.
\end{proof}

At the equalizer's output we get a corrupted version of $\mathbf{x}\otimes\mathbf{i}$, and after reducing modulo $2$, we get $\mathbf{x}_{0}$ (plus folded Gaussian noise). Since $\mathbf{x}_{0}\in\mathcal{C}$, it can be decoded before preceding to detect the uncoded bits. A complete knowledge of $\mathbf{x}_{0}$ divides the constellation into two cosets, which makes the distinction between two points from the same coset easier than the distinction between two points of the full constellation, thus doubling the Euclidean minimum distance in the constellation as in Ungerboeck's set partitioning (see, e.g., \cite{Forney98}). We can therefore first decode $\mathbf{x}_0$ and then detect the uncoded bits using a slicer with double step size. For more details, see~\cite{EilatTCM,DEFID}.

\section{Discussion and Conclusions}
\label{sec:7}
We have presented a novel DFE scheme for the discrete-time linear Gaussian channel, suitable for  single-carrier transmission where channel state information is not available at the transmitter.
The scheme enables block decoding to be performed before applying the DFE when a cyclic code is used.
The channel is equalized to an impulse response that is comprised of integer coefficients only.
The performance of the proposed scheme was analyzed, and in particular an upper bound on the induced noise enhancement was derived. Several examples of ISI channels were examined and it was shown that in many scenarios the scheme achieves a rather small gap-to-capacity.
An interesting avenue for further research is to incorporate shaping into the scheme in order to make the scheme attractive at low SNR. Another interesting question to be explored, is how IF equalization can be combined with iterative equalization methods such as turbo equalization. Since the noise enhancement the FFE in IF equalization causes is always smaller (or equal) to that of an ZF-LE, combining it with an iterative equalizer may result in better convergence.


\section*{Acknowledgement}
The authors wish to thank Jiening Zhan, Bobak Nazer and Michael Gastpar for discussions
on integer-forcing equalization. The authors also thank Yuval Kochman for helpful comments.

\begin{appendix}

\section*{The MMSE-IF Scheme}
\label{sec:5}
\begin{figure*}[t]
\psfrag{m}{$m$}
\psfrag{x}{$x_k$}
\psfrag{encoder}{\small$\text{Encoder}$}
\psfrag{s1}{\small$\Sigma$}
\psfrag{d1}{$d_k$}
\psfrag{mod1}{\tiny$\bmod \Delta$}
\psfrag{H}{\small$H(D)$}
\psfrag{IF}{\small$A(D)$}
\psfrag{Xtmp}{$\bar{x}_k$}
\psfrag{s2}{\small$\Sigma$}
\psfrag{noise}{$n_k$}
\psfrag{output}{$y_k$}
\psfrag{vk}{$v_k$}
\psfrag{s3}{\small$\Sigma$}
\psfrag{d2}{$-d_k$}
\psfrag{mod2}{\tiny$\bmod \Delta$}
\psfrag{mod3}{\tiny$\bmod \Delta$}
\psfrag{I}{\small$I(D)$}
\psfrag{s4}{\small$\Sigma$}
\psfrag{wk}{$y'_k$}
\psfrag{min}{$-$}
\psfrag{decoder}{\small$\text{Decoder}$}
\psfrag{feedback}{\small$I(D)-1$}
\psfrag{est}{$\hat{x}_k,\hat{m}$}
\psfrag{mod}{\tiny$\bmod q$}
 \includegraphics[width=2\columnwidth]{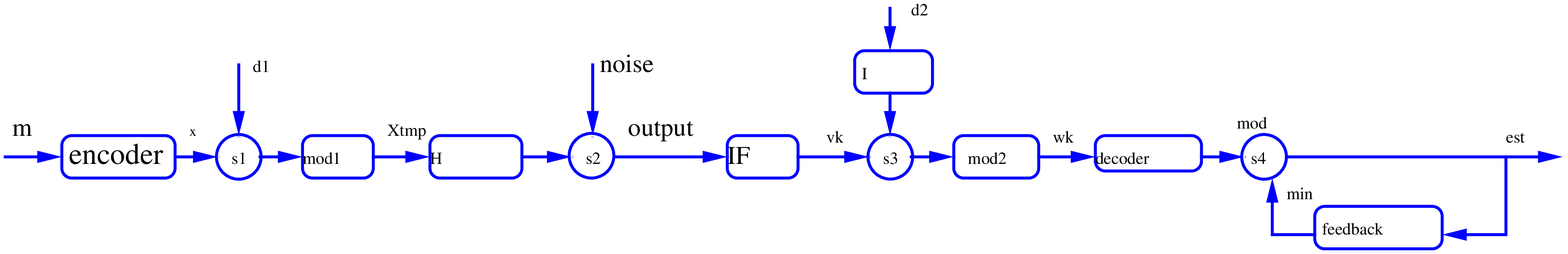}
\caption{A schematic description of a MMSE-IF system. The notation $\bmod \Delta$ denotes reducing modulo the interval $[-\Delta/2,\Delta/2)$}
 \label{MMSE-IF}
\end{figure*}
It is well known that the MMSE-DFE is strictly better than the ZF-DFE. The differences between the two are more pronounced at low SNR, and vanish as the SNR increases (for channels that satisfy the Paley-Wiener condition). We now develop an MMSE version for IF as well, which is strictly better than ZF-IF. Nevertheless, at high SNR the schemes have similar performance.

Our goal in designing the MMSE-IF equalizer is to produce an output,
\begin{align}
x_k\ast i_k+e_k=\tilde{x}_k+e_k,\nonumber
\end{align}
that maximizes the SINR, which we define as:
\begin{eqnarray*}
\text{SINR}_{\text{\rm MMSE-IF-DFE}}=\frac{\mathbf{E}\left[\tilde{x}_k^2\right]}{\mathbf{E}\left[e_k^2\right]}.
\end{eqnarray*}
Note that in fact we are interested in estimating $x'_k=x_k\otimes i_k$. However, since we transform the linear convolution performed by the channel into a cyclic one, simply by zero padding the transmitted block and ignoring the last entries of the output of the FFE (as described in Section~\ref{sec:2}), there is no difference between estimating $\tilde{x}_k$ and $x'_k$.

The output of the FFE, $v_k$, serves as an estimate for $\tilde{x}_k$ from the sequence $\{y_k\}$, and can be written as,
\begin{align*}
V(D)&=A(D)Y(D) \\
&=A(D)H(D)X(D)+A(D)N(D)\nonumber\\
&=I(D)X(D)+\left(\frac{A(D)H(D)}{I(D)}-1\right)I(D)X(D)+Z(D)\nonumber\\
&=\tilde{X}(D)+\left(G(D)-1\right)\tilde{X}(D)+Z(D),
\end{align*}
where $Z(D)$ is the filtered Gaussian noise, and $G(D)$ is some linear time-invariant filter.
Since we want our estimator to be unbiased (see \cite{Cioffi95}), we require that $G(D)$ be monic (i.e, $g_0=1$). In the time domain we have,
\begin{align}
v_k &=\tilde{x}_k+\sum_{l\neq 0}g_l\tilde{x}_{k-l}+z_k \nonumber \\
    &=\tilde{x}_k+e_k.
\label{filtOutput}
\end{align}
The derivation of the optimal MMSE-IF filter is reminiscent of finding the optimal MMSE-FFE in the classic DFE scheme \cite{Cioffi95}. However, there is one difference. In the classic DFE scheme the objective is to estimate the data symbols which are uncorrelated. Thus, the estimated symbol $x_k$ and the residual noise $e_k$ (which is composed of data symbols from different time instances and Gaussian noise) are also uncorrelated. In the MMSE-IF scheme, we estimate an (integral) {\em linear combination} of the data symbols. For this reason, different samples of $\tilde{x}_k$ are correlated and therefore $\tilde{x}_k$ and $e_k$ are correlated as well. We would like to be able to follow the derivation of the optimal MMSE-FFE \cite{Cioffi95}. To this end, we may remove the correlation between $\tilde{x}_k$ and $e_k$ by dithering as we show in Appendix~\ref{subsec:5}.

As shown in \cite{Cioffi95}, if the output of an equalizer is composed of signal and noise which are uncorrelated, maximization of the SINR is equivalent to minimization of the MSE. This implies that the optimal choice for the front-end linear filter is given by the Wiener filter,
\begin{eqnarray}
A(D)=b_0\frac{I(D)H^*(D^{-*})}{H(D)H(D^{-*})+\frac{1}{\sigma_x^2}}.
\label{IF-MMSE}
\end{eqnarray}
The role of the factor $b_0$ in (\ref{IF-MMSE}) is to guarantee that the estimator is unbiased, i.e., to guarantee that $g_0=1$. Let $S_{ee}(e^{j\omega})$ be the Power Spectral Density (PSD) of the residual noise (ignoring the bias-removal term $b_0$).
%
The filter $I(D)$ should be chosen such that,
\begin{align*}
\mathbf{E}\left[e_k^2\right]&=\frac{1}{2\pi}\int_{-\pi}^{\pi}S_{ee}(e^{j\omega})d\omega \\
&=\frac{1}{2\pi}\int_{-\pi}^{\pi}\frac{|I(e^{j\omega})|^2}{|H(e^{jw})|^2+\frac{1}{\sigma_x^2}}d\omega
\end{align*}
is minimized. Following the derivation of Section~\ref{sec:3}, we define,
\begin{eqnarray*}
\tilde{{\text{K}}}_{\text{MMSE},n}=\left[\begin{array}{ccccc}
\tilde{k}_{0} & \tilde{k}_{-1} & \tilde{k}_{-2} \ldots & \tilde{k}_{-(n-1)}\\
\tilde{k}_{1} & \tilde{k}_{0} & \tilde{k}_{-1} \ldots & \tilde{k}_{-(n-2)}\\
\vdots & \vdots & \vdots \ddots\\
\tilde{k}_{n-1} & \tilde{k}_{n-2} & \tilde{k}_{n-3} \ldots & \tilde{k}_{0}
\end{array}\right],
\end{eqnarray*}
where
\begin{eqnarray*}
\tilde{k}_{m}=\frac{1}{2\pi}\int_{-\pi}^{\pi}\frac{1}{|{H({e^{{j\omega}}})}|^{2}+\frac{1}{\sigma_x^2}}{e^{-jm\omega}}d\omega
\end{eqnarray*}
is the autocorrelation sequence corresponding to,
\begin{align}
\tilde{K}(D)=\frac{1}{H(D)H^*(D^{-*})+\frac{1}{\sigma_x^2}}.\nonumber
\end{align}
Let $\tilde{\text{F}}_{\text{MMSE},n}$ be a matrix satisfying,
\begin{eqnarray}
\tilde{\text{K}}_{\text{MMSE},n}=\tilde{\text{F}}_{\text{MMSE},n}^T\tilde{\text{F}}_{\text{MMSE},n}.\nonumber
\end{eqnarray}
The optimal choice of $I(D)$ is given by finding the shortest vector in the lattice $\tilde{\text{F}}_{\text{MMSE},n}$, which can in practice be approximated by using the LLL algorithm as was explained in Section~\ref{sec:3}. Note that the choice of $I(D)$ is now SNR dependant.

\subsection{Removing Correlation by Dithering}
\label{subsec:5}
In order to remove the correlation between $\tilde{x}_k$ and $e_k$ in (\ref{filtOutput}), we use a pseudo-random dither $d_k$ which is uniformly distributed over the modulo interval $[-\Delta/2,\Delta/2)$, and is assumed to be known to both the transmitter and the receiver. As we recall, in the case of $q$-PAM transmission, i.e., the data symbols are $$x_k\in{c\sqrt{\text{SNR}}}\left\{\frac{-(q-1)}{2q},\frac{-(q-3)}{2q},\ldots,\frac{q-1}{2q}\right\},$$ the modulo interval is simply $\left[-c\sqrt{\text{SNR}}/2,+c\sqrt{\text{SNR}}/2\right)$.  The transmitted sequence is
$$\bar{x}_k=\left(x_k+d_k\right)\bmod [-\Delta/2,\Delta/2),$$ which is uniformly distributed over $[-\Delta/2,\Delta/2)$, and statistically independent of $x_k$. Nevertheless, we still seek an estimator for $\tilde{x}_k=x_k\ast i_k$.

Define $\tilde{\bar{x}}_k$ as,
\begin{align}
\tilde{\bar{x}}_k=\bar{x}_k\ast i_k. \nonumber
\end{align}
The receiver's front-end filter is the optimal linear MMSE estimator for $\tilde{\bar{x}}_k$ from the sequence $\{y_k\}$, given by,
\begin{eqnarray}
A(D)=\tilde{b}_0\frac{I(D)H^*(D^{-*})}{H(D)H(D^{-*})+\frac{1}{\sigma_{\bar{x}^2}}},\nonumber
\end{eqnarray}
where $\tilde{b}_0$ is a bias-removal term, serving the same purpose as $b_0$ in (\ref{IF-MMSE}).
The receiver then computes,
\begin{align}
y'_k&=\left(v_k-d_k\ast i_k\right)\bmod [-\Delta/2,\Delta/2)\nonumber\\
&=\left(\tilde{\bar{x}}_k+\sum_{l\neq 0}g_l\tilde{\bar{x}}_{k-l}+z_k-d_k\ast i_k\right)\bmod[-\Delta/2,\Delta/2)\nonumber\\
&=\bigg((x_k*i_k)\bmod[-\Delta/2,\Delta/2)\nonumber\\
&+(d_k*i_k)\bmod[-\Delta/2,\Delta/2)-d_k\ast i_k \nonumber \\
& \ +\sum_{l\neq 0}g_l\tilde{\bar{x}}_{k-l}+z_k\bigg)\bmod[-\Delta/2,\Delta/2)\nonumber \\
&=\left(\tilde{x}_k+\sum_{l\neq 0}g_l\tilde{\bar{x}}_{k-l}+z_k\right)\bmod[-\Delta/2,\Delta/2)\nonumber\\
&=\left(\tilde{x}_k+e_k\right)\bmod[-\Delta/2,\Delta/2)\nonumber
\end{align}
where $\tilde{x}_k$ and $e_k$ are uncorrelated.
The complete system is illustrated in Figure~\ref{MMSE-IF}.

\end{appendix}

\bibliographystyle{IEEEtran}
\bibliography{OrBib2}
\nocite{TKS02,YaoWornell,Cover}

\end{document}